\documentclass[%
 twocolumn,
 groupedaddress,
 amsmath,
 amssymb,
 aps
]{revtex4-2}

\usepackage{
    mathrsfs,
    lipsum,    
    graphicx,
    dcolumn,
    bm,
    physics,
    csvsimple,
    soul,
    xcolor
    }
\usepackage[flushleft]{threeparttable}
\usepackage[colorinlistoftodos,prependcaption,textsize=tiny]{todonotes}

\newcolumntype{d}[1]{D{.}{.}{#1}}

\newcommand{\citeref}[1]{Ref.\,\cite{#1}}
\newcommand{\citerefs}[1]{Refs.\,\cite{#1}}
\newcommand{\citeeq}[1]{Eq.\,(\ref{#1})}

\newcommand{\Htp}[0]{H$_2^+$}

\newcommand{\HDp}[0]{HD$^+$}
\newcommand*{\Xp}{X$^+ ~ ^2\Sigma_g^+$}
\newcommand*{\Ap}{A$^+ ~ ^2\Sigma_u^+$}
\newcommand*{\X}{X$^+$}
\newcommand*{\A}{A$^+$}
\newcommand{\sphT}[2]{
    \bm #1^{(#2)}
    }
\newcommand{\threeJ}[6]{
    \left( 
        \begin{array}{ccc}
            #1 & #2 & #3 \\[2.5pt]
            #4 & #5 & #6
        \end{array} 
    \right)
}
\newcommand{\sixJ}[6]{
    \left\{ 
        \begin{array}{ccc}
            #1 & #2 & #3 \\[2.5pt]
            #4 & #5 & #6 
        \end{array} 
    \right\}
}

\begin{document}


\title{Black-body radiation induced photodissociation and population redistribution of weakly bound states in \Htp}

\author{A.\,D. Ochoa Franco}
\affiliation{Department of Physics and Astronomy, LaserLaB, Vrije Universiteit Amsterdam\\
De Boelelaan 1085, 1081 HV Amsterdam, The Netherlands}
\author{M. Beyer}
\affiliation{Department of Physics and Astronomy, LaserLaB, Vrije Universiteit Amsterdam\\
De Boelelaan 1085, 1081 HV Amsterdam, The Netherlands} 

\date{\today}

\begin{abstract}
Molecular hydrogen ions in weakly bound states close to the first dissociation threshold are attractive quantum sensors for measuring the proton-to-electron mass ratio and hyperfine-induced ortho-para mixing. The experimental accuracy of previous spectroscopic studies relying on fast ion beams could be improved by using state-of-the-art ion trap setups. With the electric dipole moment vanishing in \Htp\, and preventing fast spontaneous emission, radiative lifetimes of the order of weeks are found. We include the effect of black-body radiation that can lead to photodissociation and rovibronic state redistribution to obtain effective lifetimes for trapped ion experiments.
Rate coefficients for bound-bound and bound-continuum processes were calculated using adiabatic nuclear wave functions and nonadiabatic energies, including relativistic and radiative corrections. Effective lifetimes for the weakly bound states were obtained by solving a rate equation model and lifetimes in the range of 4 to 523~ms and $>$215~ms were found at room temperature and liquid nitrogen temperature, respectively. Black-body induced photodissociation was identified as the lifetime-limiting effect, which guarantees the purity of state-selectively generated molecular ion ensembles. The role of hyperfine-induced $g/u$-mixing, which allows pure rovibrational transitions, was found to be negligible.  
\end{abstract}

\maketitle


\section{Introduction}

 The hydrogen molecular ion (HMI) is the simplest molecular system and can be used to test quantum electrodynamics and to determine fundamental constants by comparing experimental and theoretical transition frequencies \cite{Alighanbari2020a,patra2020a}. 
While ab initio theory reached a level of $10^{-11}$ relative accuracy for \Htp\, and \HDp~\cite{Korobov2014b, Korobov2017a}, a comparable experimental accuracy has only been achieved for the \HDp~isotopologue \cite{koelemeij2007a,Alighanbari2020a,patra2020a}. 
In these experiments, rovibrationally-cold \HDp~ions are held in a radio-frequency trap and sympathetically cooled to reach the Lamb-Dicke regime in order to suppress Doppler effects during the spectroscopic interrogation.

Cooling the rovibrational degrees of freedom of the hot \HDp~ions produced by electron bombardment relies on spontaneous emission; a process that is only possible in the heteronuclear isotopologues, because of the electric dipole moment originating from the mass and charge asymmetry \cite{Bunker1974a}.  
The absence of that very dipole moment for the homonuclear isotopologues has far reaching consequences regarding the ion production, as well as the multitude of strong transitions available for spectroscopic studies: 
(i) with radiative lifetimes of the order of weeks \cite{peek1979a}, any rovibrational state distribution produced during electron bombardment will be conserved, drastically reducing the number of available ions participating in a single spectroscopic transition, and 
(ii) strong electric-dipole allowed transitions exist only between different electronic states.

Carrington and coworkers succeeded in measuring rovibronic transitions between weakly bound states of the ground (\Xp) and first excited (\Ap) electronic states of \Htp\, just below the H(1s) + H$^+$ dissociation threshold \cite{carrington1989a,carrington1989b,carrington1993a,carrington1993b,carrington1993c}. 
A fast ion beam with high current was used to compensate for the small number of ions per quantum state and having the transitions frequencies in the microwave range helped limiting the Doppler broadening. 
To avoid interaction broadening, the ion beam was directed through a sufficiently long microwave waveguide and the observation of forward and back-reflected modes allowed the cancellation of the first-order Doppler shifts. This resulted in a FWHM of 0.6~MHz and an absolute accuracy of the order of 0.5~MHz (relative accuracy $10^{-5}$).

In a new generation of measurements we aim at improving the accuracy of the transition frequencies between the weakly bound states, which have been found to show an enhanced sensitivity on the proton-to-electron mass ratio \cite{augustovivcova2014a}, by employing a similar ion trap setup as used in \citeref{patra2020a}. 
The suppression of Doppler-related effects and the careful control of magnetic fields over a small trap volume have been shown to allow to reach Hertz-level accuracy \cite{menasian1974a}.

This leaves the question of how to increase the population in individual quantum levels of the HMI, preferably creating the ions selectively in a single weakly bound rovibronic level.
Such a state-selected ion generation can be achieved using mass-analyzed threshold ionization (MATI) \cite{zhu1991a}: this involves 
photoexcitation with a laser, slightly red-detuned from the targeted level in the ion $E^+(v', N')$, which will lead via direct ionization and auto-ionization to ions in lower lying states $E^+(v \ne v', N \ne N')$ as well as high-$n$ Rydberg states converging to the $E^+(v', N')$ threshold. 
The \emph{prompt} ions in unwanted states can be spatially separated from the neutral molecules in the Rydberg states, which are subsequently field-ionized to selectively provide ions in the state $E^+(v', N')$ for subsequent experiments \cite{mackenzie1994a}.

In order to apply MATI for the efficient production of the weakly bound states in the vicinity of the dissociation threshold, a multi-step excitation pathway has to be employed to gradually increase the bond length from $\sim 1.4\,a_0$ (X$(v=0, N=1)$) to $\sim 23.4\,a_0$ (\X\,$(v=19, N=1)$), as shown in the photoelectron spectroscopic studies by Beyer and Merkt for H$_2$, HD and D$_2$ \cite{Beyer2016a,beyer2018a,beyer2022a}.
The crucial ingredient for the excitation of weakly bound ions, starting from the vibronic ground state of neutral molecular hydrogen, was the use of the long-range $\mathrm{H}\mathrm{\bar H}$ and $\mathrm{B}\mathrm{\bar B}$ intermediate states, first observed and characterized by W. Ubachs and his coworkers \cite{reinhold1997a,Reinhold1999a,deLange2001a}. 
These states belong to the class of \emph{ion-pair states} \cite{reinhold2005a} and are characterized by large bond lengths and mixed electronic character (regarding the orbital-angular momentum $\ell$ in a single-center description), allowing for the efficient generation of weakly bound molecular ions with $N=0-10$. 

Previous theoretical studies on the radiative lifetimes of the weakly bound states have shown lifetimes in excess of hundreds of seconds \cite{peek1979a,moss1993a}, even when including the effect of ortho-para or $g/u$-mixing due to the hyperfine structure \cite{bunker2000a}, making them very attractive for precision measurements. 
However, neither of these studies addressed the effect of photodissociation and state-redistribution induced by black-body radiation (BBR).
The peak of the black-body spectrum (BBS) at room temperature is located at around $17$~THz, overlapping with the electric-dipole spectrum of the weakly bound states and supporting the possibility of having a perceptible effect on the lifetimes and the state-distribution of the selectively prepared ions.

In the following, we present a theoretical study on the effective lifetimes of HMI in weakly bound states, taking the effects of BBR and $g/u$-mixing into account. 
Section \ref{sec:theory} shows the calculation of the Einstein coefficients for bound-bound and bound-continuum transitions, required to solve the rate equations for the HMI. Effective lifetimes and the time evolution of the state distributions are shown in section \ref{sec:results} and in section \ref{sec:discussion} the results are discussed in view of the planned precision measurements mentioned above.

\section{Theory}
\label{sec:theory}

\subsection[]{Level Structure}
\label{subsec:rovib}

\begin{figure}
\includegraphics[scale=0.925]{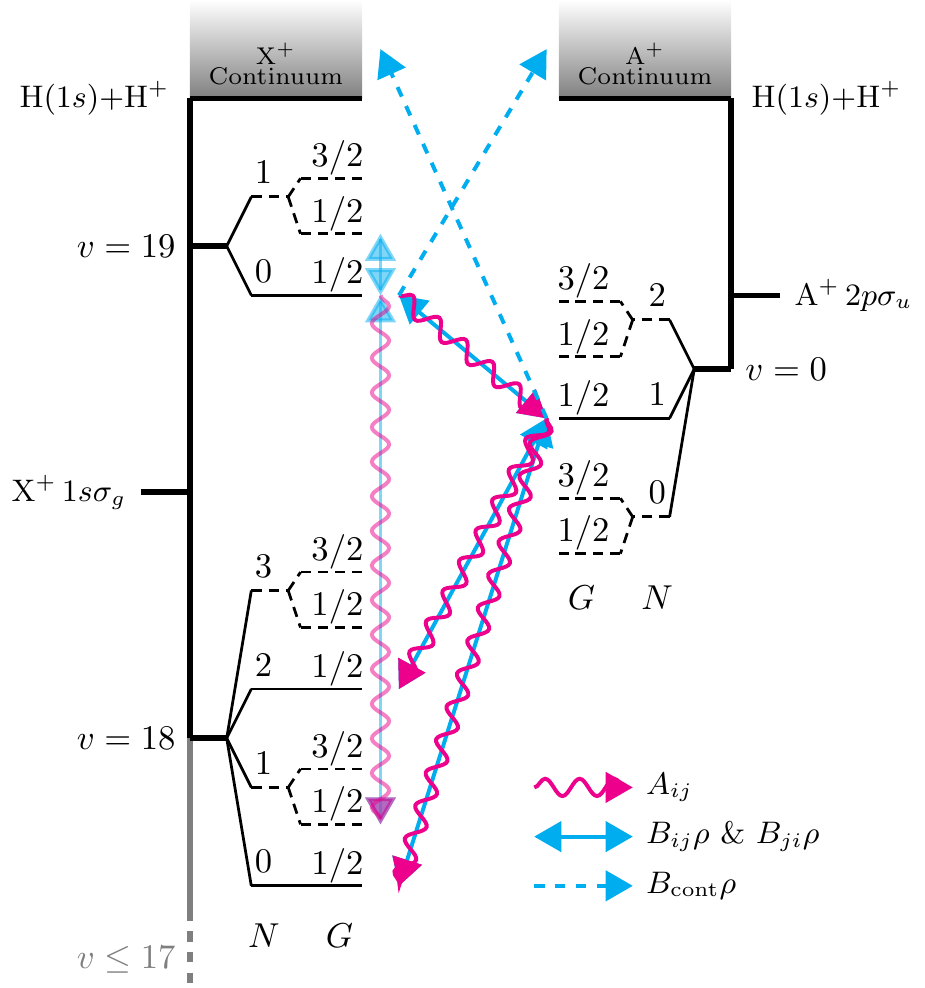}
\caption{
    Energy-level diagram of the weakly bound states of \Htp. The ground electronic state, \Xp, is depicted on the left-hand side, displaying the dissociation continuum on the top and the two highest vibrational levels. Levels with $v=13-17$ are not shown for clarity, but are taken into account in the calculation 
    On the right-hand side, the first excited electronic state, \Ap, is shown with a single vibrational level, $v=0$, and the dissociation continuum.
    For both electronic states, the rotational levels denoted with a dashed line are the ortho-levels with $I=1$. 
    The arrows represent a simplified version of the transitions present for para-\Htp\, with the whole population starting in the \X($v=19,N=0,G=1/2$) level; wavy red lines indicate spontaneous emission, the double-headed solid blue lines show stimulated emission and absorption, and the single-headed dashed blue line stands for absorption into the continuum (all states experience absorption into the continuum, but only a few are explicitly shown). 
    The forbidden transitions, allowed by g/u-mixing, are also shown.
    Neglecting spin-rotation interactions, the levels $F$ are degenerate and not shown.
}
\label{fig:energyLevels} 
\end{figure}

In this study, we consider electric-dipole transitions between rovibrational levels of the the electronic ground and excited state \Xp\, and \Ap, respectively.
To determine the transition strengths, the vibrational wave functions are obtained by solving the nuclear Schrödinger equation 
\begin{align}
\label{eq:schro}
    \left[ -\frac{1}{2\mu}\frac{\mathrm{d}^2}{\mathrm{d}R^2} + \frac{N(N+1)}{2\mu R^2} + U(R) - E_{vN} \right] \psi_{vN}(R) = 0 \,\text{,}
\end{align}
\noindent where $\mu$ is the nuclear reduced mass and $R$ is the internuclear distance. $U(R)$ is the Born-Oppenheimer (BO) potential energy curve with the adiabatic correction added and was taken from \citeref{beyer2016b}. Atomic units and the values of the fundamental constants from 2018 CODATA \cite{tiesinga2021codata} were used.
The integration is performed using the renormalized Numerov method \cite{Johnson1977} with a radial step size of $0.01a_0$, resulting in dissociation energies in agreement with the ones given in \citeref{moss1993a}. Small deviations below 1~cm$^{-1}$ originate from disregarding nonadiabatic, relativistic and QED effects in the current calculation, which tend to vanish for the dissociation energies of the weakly bound states, as the molecular corrections approach the atomic values. The rate coefficients in Section \ref{subsec:rate_eqns} are obtained using the adiabatic nuclear wave functions, but employing the exact nonadiabatic energies (including relativistic and radiative corrections) from \citeref{moss1993a}.

The \Xp\, state has 19 vibrational states with at least four rotational levels, $0\le N \le 3$, and $v=19$, which has only two rotational levels, $N=0$ and $N=1$. The repulsive \Ap\, state supports one vibrational state, $v=0$, with three rotational levels, $0\le N \le 2$. A $v=1$ halo state exists when the hyperfine structure is neglected \cite{carbonell2003a,beyer2018a}, but was disregarded in the current work because of an insignificant Franck-Condon overlap. The relevant weakly bound states are depicted in the energy-level diagram in Fig.\,\ref{fig:energyLevels}, where it can be seen that the \X\,$(19,N = 0,1)$ levels lie above the \A\,$(0)$ manifold, allowing for spontaneous emission.
Rovibrational states of \X\, with $13 \leq v \leq 19$ and $0 \leq N \leq 3$ are considered, with the range of $v$ based on the value of the Franck-Condon factor, which rapidly decreases from $\sim0.5$ for \A\,$-\,$\X\,$(19)$ to $\sim10^{-8}$ for \A\,$-\,$\X\,$(13)$.

The dissociation continua of the \X\, and \A\, state, indicated as a shaded area at the top of Fig.\,\ref{fig:energyLevels}, play a crucial role in the present calculation.
The continuum wave functions are obtained using outward integration of \citeeq{eq:schro} and are energy normalized by scaling their amplitude to $\sqrt(2\mu/\pi k)$ with $k=\sqrt(2\mu(\epsilon-U(\infty)))$, and $\epsilon$ being  the kinetic energy of the continuum state \cite{merzbacher1961a}.
The energy-normalized continuum wave functions have dimensions of [(Length Energy)$^{-1/2}$].
To adequately account for the photodissociation of the \X\,$(v,2)$ and \X\,$(v,3)$ states, the continuum of \A\, is taken into account with $N\leq 4$.
\\

The hyperfine structure of the rovibrational levels in \Htp\, is described using Hund's case $b_{\beta S}$ with the coupling scheme:
\begin{align}
    \bm I = &\, \bm I_1 + \bm I_2 \,\text{,} \notag\\
    \bm G = &\, \bm S + \bm I \,\text{,} \notag\\
    \bm F = &\, \bm G + \bm N \,\text{.} \notag
\end{align}
The coupling of $\bm S$ and $\bm I$ to form $\bm G$ is caused by the Fermi contact interaction,
\begin{align}
\label{eq:H_F}
    H_{F} = &\, b_F(v,N) \, \sphT{S}{1} \cdot \sphT{I}{1} \,\text{,} 
\end{align}
\noindent which is the leading term in the hyperfine Hamiltonian with $b_F(v,N)$ of the order of $\sim700$~MHz. The coupling constants were computed using the adiabatic nuclear wave functions and the electron densities from \citeref{beyer2018a} (see Tables \ref{table:bF} and \ref{table:bF_A} in Appendix\,\ref{sec:bF}). For \X\,$(19,1)$ we find excellent agreement with the experimentally determined value from \citeref{carrington1993a}.

The two nuclear-spin modifications para, with $I=0$, and ortho, with $I=1$, combine with even (odd) and odd (even) rotational levels of the \Xp (\Ap) state, respectively, as imposed by the Pauli principle. The rotational levels of ortho-\Htp\, are indicated using dashed lines in Fig.\,\ref{fig:energyLevels}.
The Fermi contact interaction in \citeeq{eq:H_F} splits each rotational level in ortho-\Htp\, in a $G=1/2$ and $G=3/2$ component, whereas no splitting appears for para-\Htp\, with $G=S=1/2$. The couplings of the electron and nuclear spins to the molecular rotation are weak and were neglected, which leads to a degeneracy of all $F$ components for a given $(N, G)$ pair.

Close to the dissociation limit the hyperfine interaction becomes comparable to the splitting of the \emph{gerade} and \emph{ungerade} electronic states, which leads to $g/u$-mixing (or ortho-para mixing) of levels with $G=1/2$ and the same value of $N$ \cite{beyer2018b}.\\

\subsection[]{Line strengths}
\label{subsec:mue}

To calculate the population redistribution induced by BBR, we evaluate the line strength for the electric-dipole transition from lower state $\ket{\eta \Lambda; v; N S I G F}$ to upper state $\ket{\eta' \Lambda'; (v'/\epsilon); N' S I G' F'}$, defined as \cite{larsson1983a}
\begin{widetext}
\begin{align} 
\label{eq:S_hfs}
    S_{\eta',v',F';\eta,v,F} = & \, \sum_{MM'} |\bra{\eta' \Lambda'; (v'/\epsilon); N' G' F' M'} 
        \, \sphT{\mu}{1}_{e}
        \ket{\eta \Lambda; v; N G F M}|^2 \nonumber \\
        \, = & \,  \mathscr{S}_{\Lambda'N'G'F',\,\Lambda NGF}
        \, |\bra{\eta'\Lambda'; (v'/\epsilon) N'} \sphT{\mu}{1}_{e,q} \ket{\eta \Lambda; v N}|^2  \, \text{,}
\end{align}
\noindent which was expressed in a rotational and vibronic part in the second line, with the electric-dipole transition moment $\sphT{\mu}{1}_{e}$. The quantum numbers $v'$ or $\epsilon$ indicate if the upper state is bound or a continuum state. For clarity, the quantum numbers $S=1/2$ and $I=0,1$ are not explicitly written and the vibrational quantum number $v$ is replaced with $\epsilon$ for continuum states. After transforming to the molecular-fixed coordinate system, the rotational integral can be obtained by making use of the following matrix element \cite{brown2003a,zare1988a}

\begin{equation}
\label{eq:HLF_hfs}
\begin{split} 
    &| \bra{ \Lambda' N' G' F' M'} \bm{D}_{pq}^{(1)*} \ket{ \Lambda N G F M}|^2 \\
        &\quad = \delta_{GG'} (2F+1)(2F'+1)(2N+1)(2N'+1)
            \threeJ{N'}{1}{N}{-\Lambda'}{q}{\Lambda}^2 \sixJ{N'}{F'}{G}{F}{N}{1}^2 \, \threeJ{F'}{1}{F}{-M'}{p}{M}^2  \, \text{.} 
\end{split}
\end{equation}

\end{widetext}

For $\Sigma-\Sigma$ transitions, $q=0$, and these matrix elements vanish unless $\Delta G = 0$ and $\Delta N = \pm 1$.
For isotropic excitation (BBR is unpolarized) and emission, \citeeq{eq:HLF_hfs} is summed over $p, M$ and $M'$, and the Hönl-London factor $\mathscr{S}_{\Lambda'N'G'F',\,\Lambda NGF}$ including hyperfine-structure is obtained as:
\begin{align} \label{eq:HLF_sums}
    \mathscr{S}_{\Lambda'N'G'F',\Lambda NGF} &\, \nonumber\\
            = \delta_{GG'}&\,(2F+1)(2F'+1) (2N+1)(2N'+1) \nonumber\\
                &\, \times \threeJ{N'}{1}{N}{-\Lambda'}{q}{\Lambda}^2 \sixJ{N'}{F'}{G}{F}{N}{1}^2 \,\text{.}
\end{align}

In the following, the Hönl-London factor for a rotational line without and with spin is considered.
\noindent Neglecting spin, with $(S=0,I=0) \rightarrow G=0$ and $F=N$, \citeeq{eq:HLF_sums} simplifies to the more common expression
\begin{align} 
\label{eq:Xi}
    \mathscr{S}_{\Lambda'N',\,\Lambda N} =
        (2N+1)(2N'+1)
        \threeJ{N'}{1}{N}{-\Lambda'}{q}{\Lambda}^2 \, \text{.}
\end{align}
For the general case in \citeeq{eq:HLF_sums}, we establish the sum rule
\begin{align} \label{eq:HLF_band}
    \mathscr{S}_{\Lambda'N'G,\,\Lambda NG} \equiv &\, \sum_{FF'} \mathscr{S}_{\Lambda'N'GF',\,\Lambda NGF} \nonumber\\
                                    = &\, (2G+1)\,  \mathscr{S}_{\Lambda'N',\,\Lambda N} \, \text{,}
\end{align}
using the orthogonality of the 6-$j$ symbols \cite{zare1988a,watson2008a}. Notice, that the right-hand side of this equation depends on spin only through the factor of $(2G+1)$. When $I=0$, $(2G+1)\rightarrow(2S+1)$, and the spin-independent Hönl-London factor is recovered, multiplied with the electron spin degeneracy factor $(2S+1)$ \cite{lefebvre2004a,watson2008a}.

The vibronic matrix elements of $\sphT{\mu}{1}_{e,q}$ were calculated using the adiabatic nuclear wave functions from \citeeq{eq:schro} and the electric-dipole transition moment, $\mu_{\text{X}^+\text{-A}^+}(R)$ from \citeref{beyer2016b}.
For bound-bound transitions we have
\begin{align} \label{}
    \mu_{v'N',vN} \equiv &\, \bra{\text{A}^+,v' N'} \sphT{\mu}{1}_{e,0} \ket{\text{X}^+,vN}  \nonumber \\
    = \int \mathrm{d}R &\,\, \psi_{\text{A}^+ v'N'} \,\, \mu_{\text{X}^+\text{-A}^+} \,\, \psi_{\text{X}^+ vN} \, \text{,}
\end{align}
and the transition moments for pure rovibrational transitions due to $g/u$-mixing were taken from \citeref{bunker2000a}.

For bound-continuum transitions to a dissociative state with kinetic energy $\epsilon$, for instance \X\,$(v,N)\,-\,$\A\,$(\epsilon,N')$, the matrix elements of $\sphT{\mu}{1}_{e,q}$ are given by
\begin{align} \label{eq:mue_v}
    \mu_{N',vN}(E_\gamma) \equiv &\, \bra{\text{A}^+,\epsilon N'} \sphT{\mu}{1}_{e,0} \ket{\text{X}^+,vN} \, \nonumber \\
    = \int \mathrm{d}R &\,\, \psi_{\text{A}^+ \epsilon N'} \,\, \mu_{\text{X}^+\text{-A}^+} \,\, \psi_{\text{X}^+ v N} \, \text{,}
\end{align}
where $\mu_{N',vN}(E_\gamma)$ are continuous functions of the photon energy,
 \begin{align} \label{eq:mue_k}
    E_\gamma = \epsilon - E_{vN} \,\text{,}
\end{align}
and $E_{vN}<0$ is the binding energy of the bound state.
Notice that since the continuum wave functions are energy-normalized, this transition moment has units of $(e a_0)/E_H^{1/2}$.

Finally, \citeeq{eq:S_hfs} can be written for our specific case as
\begin{align} \label{}
\label{eq:Sij}
    S_{\eta'v'N'G, \, \eta vNG} = (2G+1)\, \mathscr{S}_{N',\, N} |\mu_{\eta'v'N',\eta vN}|^2  \, \text{.}
\end{align}

The photodissociation cross section (PDCS) in atomic units ($a_0^2$) is given by \cite{carrington1993a,lefebvre2004a}
\begin{align} 
\label{eq:cs}
    \sigma_{\eta v N}(E_\gamma) = 
        \frac{4}{3} \pi^2 \, \sum_{N'} \alpha E_\gamma
        \frac{\mathscr{S}_{N',N}\,|\mu_{N',vN}(E_\gamma)|^2}{(2N+1) } \, \text{,}
 \end{align}
and is obtained by summing over all $(F,F')$ and $N'$. Using \citeeq{eq:HLF_band}, the total state degeneracy factor $(2F+1)$ in the denominator simplifies to $(2N+1)$.

\subsection[]{Einstein coefficients and rate equations}
\label{subsec:rate_eqns}

To determine the time evolution of an ion ensemble prepared in a single initial state, we solve the rate equations of the form
\begin{align}
\label{eq:rate_eqns}
    \frac{\mathrm{d}\mathrm{N}_j(t)}{\mathrm{d}t} = 
        & \, \sum_{i>j} A_{ji} \mathrm{N}_i(t) - \sum_{i<j} A_{ij} \mathrm{N}_j(t) \\
        & + \sum_{i \neq j} \bar B_{ji} \mathrm{N}_i(t) - \sum_{i \neq j} \bar B_{ij}\mathrm{N}_j(t) - \bar B_{\text{cont},j} \mathrm{N}_j(t) \, \text{,}\nonumber 
\end{align}
\noindent where the notation has been shortened, where appropriate, such that the indices $i$ and $j$ stand for any state $\ket{\eta vN}$, $\mathrm{N}_j$ is the number of ions in the state $j$, $A_{ij}$ is the rate of spontaneous emission for the transition $j \rightarrow i$, $\bar B_{ij}$ is the rate of stimulated emission, $\bar B_{ji}$ is the rate of absorption, and $\bar B_{\text{cont}}$ is the rate of absorption to the continuum.

The rate of spontaneous emission in atomic units ($E_H/\hbar$) is given by
\begin{align}
\label{eq:Aij}
    A_{ij} =\frac{4}{3} \left( \alpha E_\gamma \right)^3 \frac{S_{ij}}{g_j} \text{,} 
\end{align}
\noindent  where $g_j$ is the degeneracy of the initial state. It can be shown that this equation determines the decay rate of, both, a rotational line without spin using the Hönl-London factor from \citeeq{eq:Xi} with $g_j=2N+1$, and a hyperfine level $F$ using \citeeq{eq:HLF_hfs} with $g_j=2F+1$ \cite{whiting1972a}.
However, to obtain the decay rate for a rotational line with spin it is necessary to sum over the fine or hyperfine structure levels $F$ and $F'$ \cite{whiting1972a,larsson1983a,thorne1999a}, that is,
\begin{align}
\label{eq:A_sum}
    A_{v'N',vN} = &\, \frac{ \sum_{FF'} A_{v'N'GF',vNGF} }{ \sum_F (2F+1) } \nonumber\\
        = &\, \frac{ \sum_{FF'} \mathscr{S}_{\Lambda'N'GF',\Lambda NGF} |\mu_{v'N',vN}|^2 }{ \sum_F (2F+1) } \,\text{,}
\end{align}
\noindent where $\sum_F (2F+1) = (2G+1)(2N+1)$ \cite{watson2008a}.
This can be simplified to
\begin{align}
\label{eq:Aij_}
    A_{v'N',vN} = &\, \frac{4}{3} \left( \alpha E_\gamma \right)^3 
        \frac{\mathscr{S}_{N',N}}{(2N+1)} |\mu_{v'N',vN}|^2 \,\text{.}
\end{align}
Similar to the PDCS the dependence on $G$ cancels out, so that different hyperfine levels $G$ from a rovibrational state $(vN)$ decay at the same rate to another state $(v'N')$.
The combination of Hönl-London factor and initial state degeneracy in \citeeq{eq:Aij} for different cases can be summarized as follows:
\begin{equation}
    \begin{split}
        \mathscr{S}_{ij} = 
        \begin{cases}
          \text{Eq.\,}(\ref{eq:Xi})\\
          \text{Eq.\,}(\ref{eq:HLF_band}), G \rightarrow S\\
          \text{Eq.\,}(\ref{eq:HLF_band})\\ 
          \text{Eq.\,}(\ref{eq:HLF_sums})\\
          \text{Eq.\,}(\ref{eq:HLF_hfs})
        \end{cases}
    \end{split}
    \begin{split}
        g_j =
        \begin{cases}
          (2N+1)\\
          (2N+1)(2S+1)\\
          (2N+1)(2G+1) \,\text{,}\\
          (2F+1)\\
          1
        \end{cases}    
    \end{split}     
\end{equation}
\noindent where the first line is for a spin-less rotational line, the second line is for a rotational line with electronic spin and $I=0$, the third line is for a rotational line with nuclear spin, the fourth line is for a hyperfine level $F$, and the last line is for a Zeeman component of the hyperfine structure.

The rate of  stimulated emission  $\bar B_{ij}$ in \citeeq{eq:rate_eqns} has the same units as $A_{ij}$ and is related to the Einstein $B_{ij}$ coefficient through the photon density, $\rho(T,E_\gamma)$, that is,
\begin{align}
\label{eq:Bbar}
    \bar B_{ij} \equiv B_{ij}\,\rho(T,E_\gamma) \,\text{,}   
\end{align}

\noindent where $B_{ij}$ is given in atomic units $a_0^3 E_H/\hbar$ by
\begin{align}
\label{eq:Bij}
    B_{ij} = &\,  \frac{4}{3}\pi^2 \frac{S_{ij}}{g_j}\quad \text{and} 
\end{align}
\begin{align}
\label{eq:BBS}
    \rho(T,E_{\gamma}) = \frac{1}{\pi^2} 
        \left( \alpha E_{\gamma} \right)^3 \frac{1}{\exp(E_{\gamma}/k_B T)-1} \, \text{,}
\end{align}
with atomic units $1/a_0^3$.
\noindent For absorption, the rate coefficient is expressed as
\begin{align}
\label{eq:B_abs}
    B_{ji} = (g_j/g_i)\, B_{ij} \,\text{.} 
\end{align}

The rate of absorption to the continuum -- or photodissociation rate --, $\bar B_{\text{cont}}$, was calculated using \citeeq{eq:Bbar} and \citeeq{eq:Bij} considering the $E_\gamma$-dependent electric dipole moment, as in \citeeq{eq:cs}. 
Then, we integrated with respect to $E_\gamma$ and summed over the possible $N'$ values in the continuum, to obtain
\begin{align}
\label{eq:Bk_}
    &\bar B_{\text{cont},\eta vN} \, = \\
         &\, \frac{4}{3}\pi^2 
            \sum_{N'} \int \mathrm{d}E_\gamma \frac{\mathscr{S}_{N',N}\,|\mu_{e,\eta vNN'}(E_\gamma)|^2}{(2N+1)} 
            \rho(T,E_{\gamma}) \,\text{.} \nonumber
\end{align}
\noindent This equation is equivalent to the photodissociation rate \cite{koelemeij2011a}, $\Gamma_{\text{PD}} = \int dE_\gamma \sigma(E_\gamma) I(T,E_\gamma)/E_\gamma$, where $I$ is the radiation intensity distribution.
Numerical integration of this equation is performed using Gauss-Legendre quadrature with 4000 energy steps in the range $0<E_\gamma<0.035E_H$.

\section{Results}
\label{sec:results}

\begin{figure}[!htb]
    \includegraphics[scale=0.95]{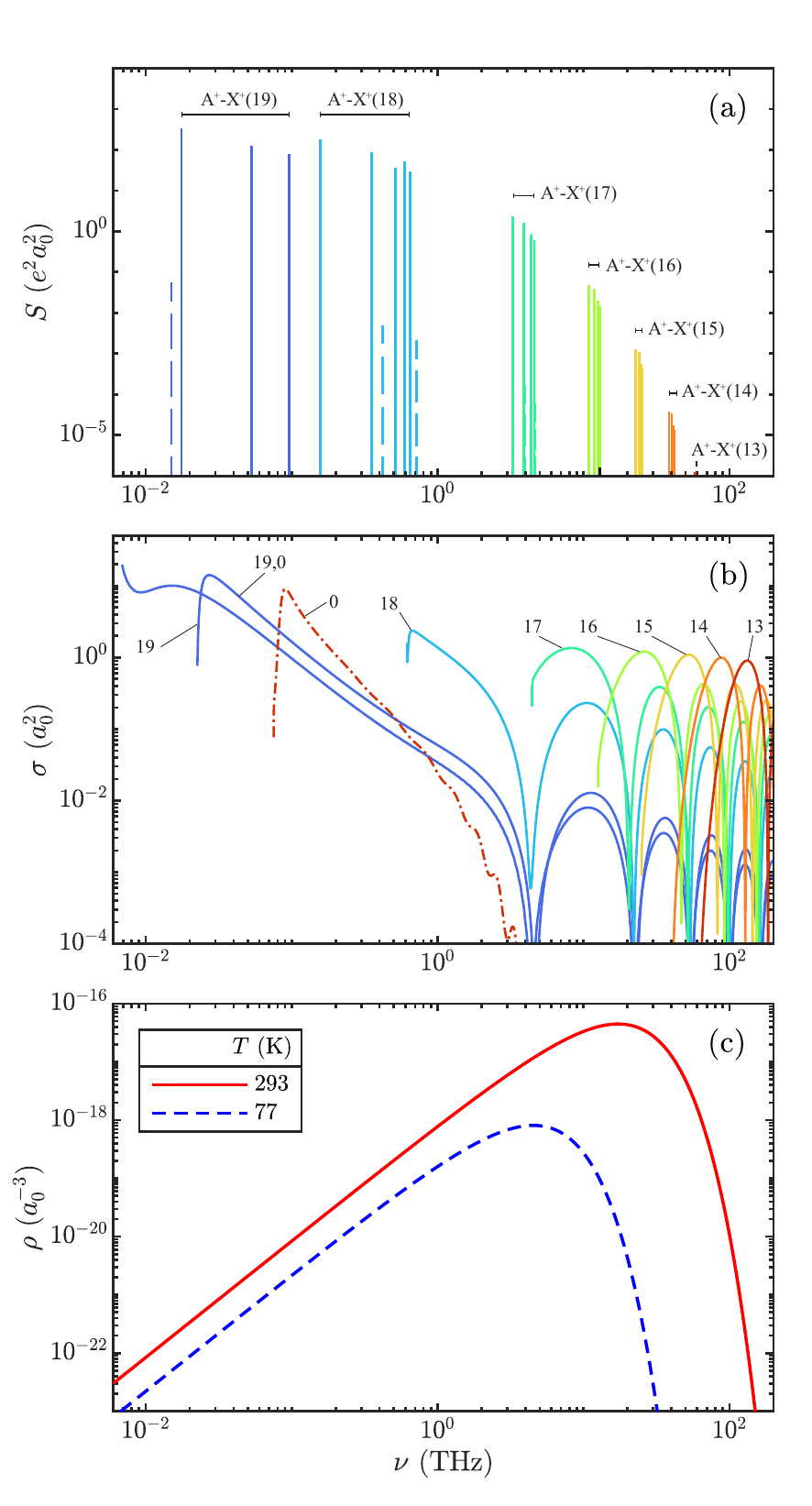}
    \caption{
        (a) Bound-bound \A\textemdash\X\, stick spectrum of \Htp, with the labels indicating the vibrational bands.
        The forbidden transitions (allowed by g/u-mixing) are indicated as dashed lines.
        (b) Photodissociation cross section for \X$(v=13-19,N=1)$ (solid line) and \A$(v=0,N=1)$ (dash-dotted line). For comparison, \X$(19,0)$ is also shown.
        (c) Black-body photon density at room (solid line) and at liquid nitrogen (dashed line) temperature. See text for details. 
    }
    \label{fig:spectra}
\end{figure}

Fig.\ref{fig:spectra}(a) displays the \A\textemdash\X\, stick spectrum with the line strengths according to \citeeq{eq:Sij}, which agree with those of \citeref{howells1991a} within 0.03\%. The transitions appear grouped in vibrational bands and spread over a frequency range from 14~GHz to 62~THz.  
In the low frequency region, the transitions involving the two highest vibrational levels of the ground state ($v=18,19$) exhibit the largest line strengths, up to 338~$e^2a_0^2$. This is caused by the favorable Franck-Condon factors and the large transition dipole moment (proportional to the internuclear distance), which is characteristic for charge-transfer transitions \cite{mulliken1939a}.
It can be noticed, that the spacing between individual rotational lines and vibrational bands become comparable when approaching the dissociation threshold, indicating the breakdown of the BO approximation and a reversal of the typical hierarchy of motion. 

The dashed lines in Fig.\ref{fig:spectra}(a) indicate the transitions, which appear because of intensity borrowing due to the $g/u$-mixing. The strongest additional line is attributed to the purely rotational X(19,0)-(19,1) transition, which was experimentally observed by Critchley and coworkers \cite{critchley2001a}.

The PDCS for \X$(v=13-19,N=1)$ and \A$(v=0,N=1)$ as obtained from \citeeq{eq:cs} is shown in Fig.\,\ref{fig:spectra}(b).
For clarity, we omitted the PDCS of the levels with $N \neq 1$, which show a similar shape. The largest $N$-dependence of the PDCS is expected for \X(19) and can be seen to be of the order of $1~a_0^2$,
by comparing the curves for \X(19,0) and \X(19,1). 
For lower vibrational states, the difference in the PDCS vanishes - in accordance with the BO approximation.
Three relevant characteristics of the PDCS can be noticed:
(i.) A threshold appears, located at the dissociation energy of the rovibrational level, which increases with decreasing $v$, (ii.) They possess a global maximum close to their threshold, and (iii.) Successive local maxima decrease in amplitude.

Employing \citeeq{eq:BBS}, Fig.\,\ref{fig:spectra}(c) illustrates the BBS photon density at room temperature (293~K) and at liquid nitrogen temperature (77~K), with maxima at 17~THz and 4~THz, respectively. A comparison of Fig.\,\ref{fig:spectra}(a-c) clearly exemplifies the importance of BBR induced processes for bound-bound and bound-continuum transitions involving the weakly bound states, with the BBS peak overlapping with the maximum of the PDCS of the levels \X\,$(v=16-17)$.

To study the time evolution of the ion's rovibronic state distribution under the influence of BBR, the rate equations in \citeeq{eq:rate_eqns} were solved for different initial conditions, corresponding to having all ions in a single rovibronic state.

\begin{figure*}
    \includegraphics[scale=0.95]{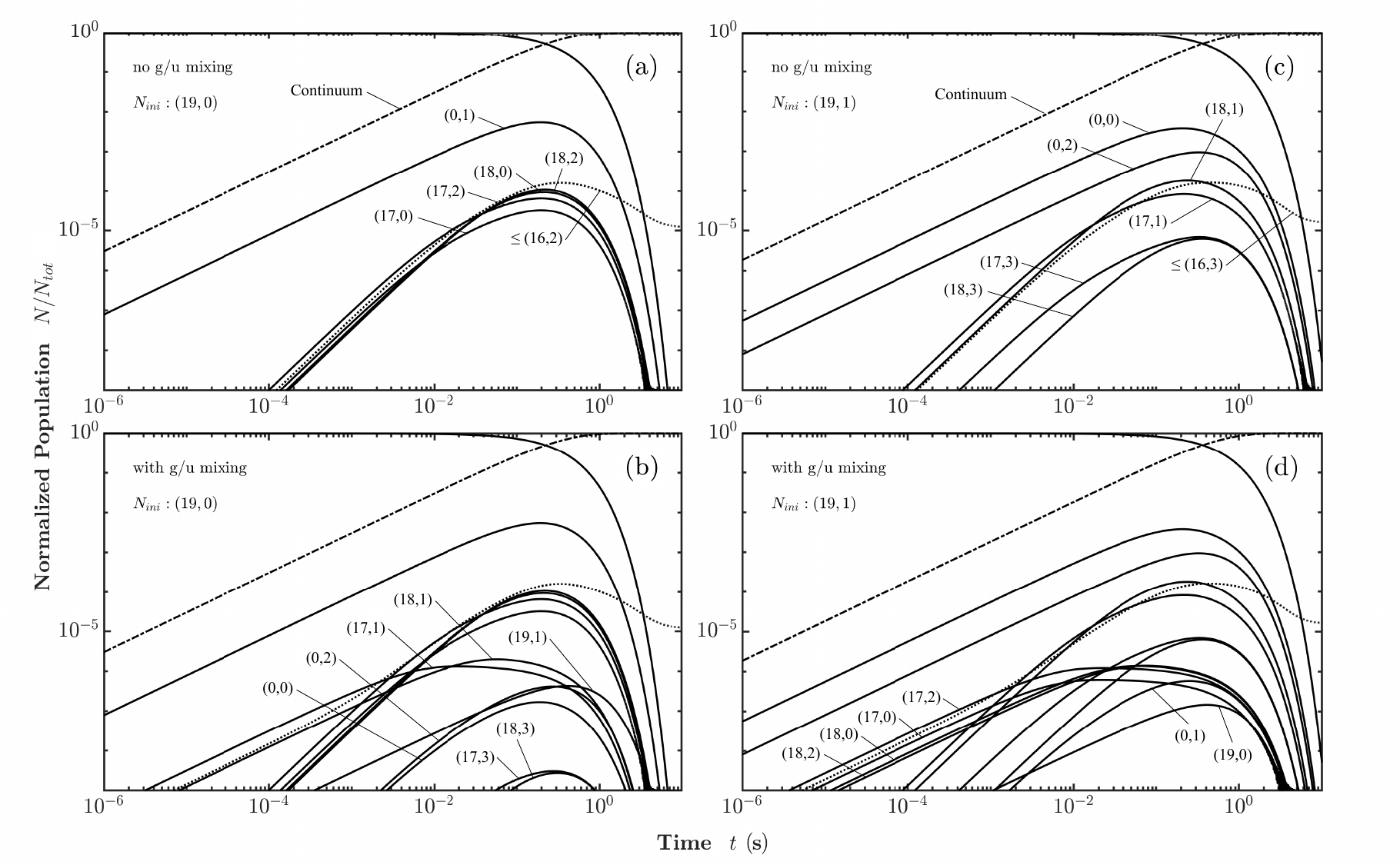}
    \caption{
    Time evolution of the rovibronic state populations at 293~K with the ions selectively prepared in:
    (a) para-\X\,$(19,0)$ with no $g/u$-mixing.
    (b) para-\X\,$(19,0)$ including $g/u$-mixing.
    Labels are added to the now-accessible ortho-states and the repeated labels from panel (a) are omitted.
    (c) ortho-\X\,$(19,1)$ without $g/u$-mixing.
    (d) ortho-\X\,$(19,1)$ with $g/u$-mixing.
    }
    \label{fig:bbr}
\end{figure*}

\begin{figure*}
    \includegraphics[scale=0.95]{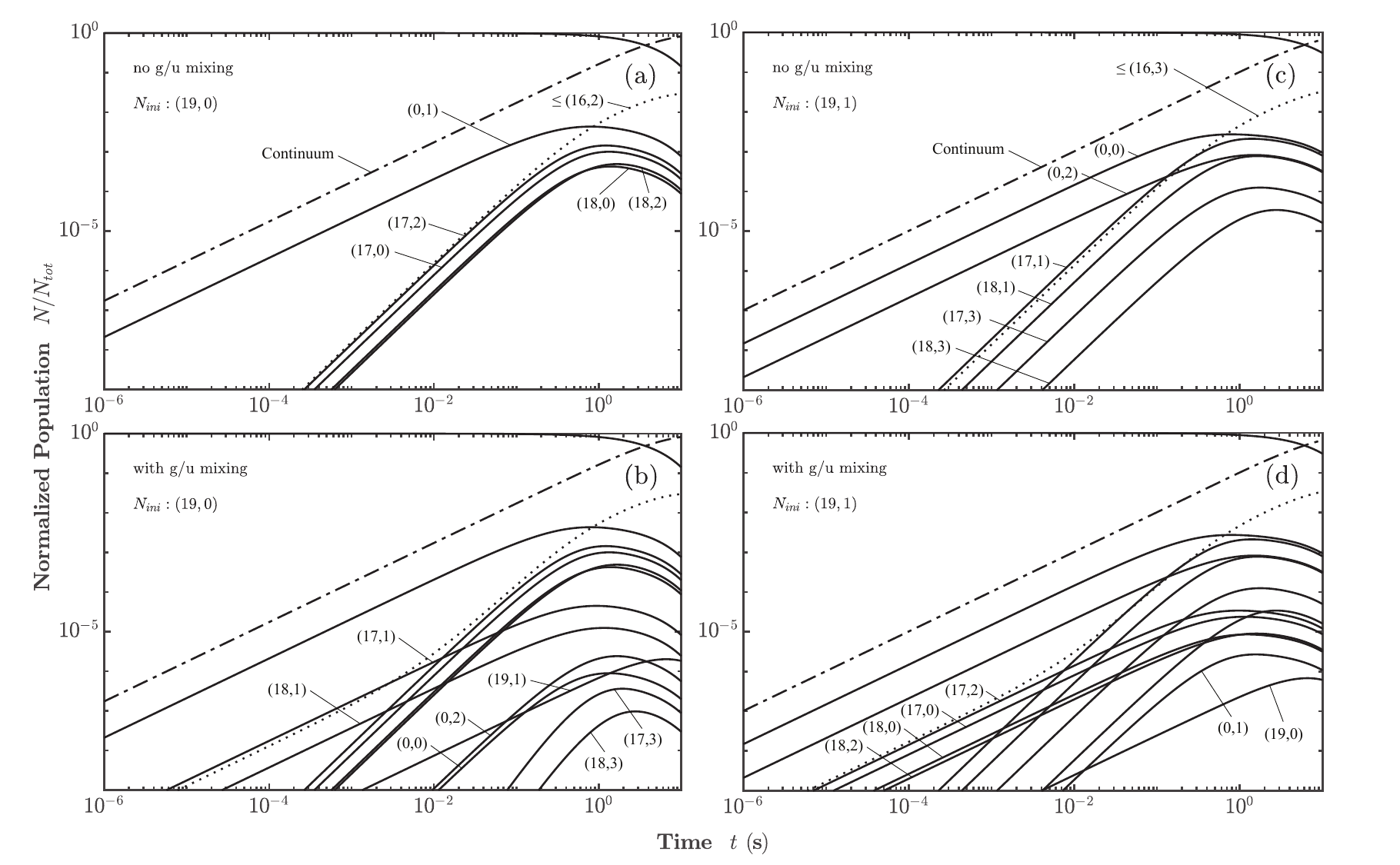}
    \caption{
    Time evolution of the rovibronic state populations at 77~K with the ions selectively prepared in:
    (a) para-\X\,$(19,0)$ with no $g/u$-mixing.
    (b) para-\X\,$(19,0)$ including $g/u$-mixing.
    Labels are added to the now-accessible ortho-states and the repeated labels from panel (a) are omitted.
    (c) ortho-\X\,$(19,1)$ without $g/u$-mixing.
    (d) ortho-\X\,$(19,1)$ with $g/u$-mixing.
    }
    \label{fig:bbr_77K}
\end{figure*}

Fig.\,\ref{fig:bbr} and Fig.\,\ref{fig:bbr_77K} display the internal state distributions as a function of time, with the population being initially in the \X(19,0) and \X(19,1) state at 293~K and 77~K, respectively. The possible excitation and decay pathways for ions initially in the \X(19,0) state are indicated in Fig.\,\ref{fig:energyLevels}, where \X$(v\le17)$ states are not shown for clarity, but are connected in the same way as \X(18). 
Solving the corresponding rate equations for $T=293$~K leads to the normalized populations as shown in Fig.\,\ref{fig:bbr}(a). As dictated by the electric-dipole selection rules, the \X(19,0) population can only be transferred to the \A(0,1) bound state, or into the \A dissociation continuum, with the latter process being much more efficient. Subsequently, the \A(0,1) state is transferred to the lower lying \X$(v\le18)$ levels via spontaneous and stimulated emission. Whereas stimulated emission is dominant at 293~K for transitions down to \X(17,2), spontaneous emission becomes the dominant process for the population of the level \X(17,0) and below. 
Numerical values of the rate coefficients can be found in the Tables \ref{table:mue_para}-\ref{table:Bk_77K} in Appendix \ref{sec:num_res}.
After a few hundred milliseconds, the population of the initially unpopulated bound states reaches a maximum, before ultimately approaching zero because of BBR induced photodissociation. A noteworthy exception are \X\, states with $v\le16$, approximately 0.001\% of the initial population, which are trapped, because the BBR does not provided enough photon flux at these high frequencies.

All states populated during the decay of \X(19,0) are para-\Htp states with $G=1/2$, unless the g/u-mixing is taken into account. The mixing of gerade and ungerade levels allows pure rovibrational transitions, which were included in the rate equations, and the results are shown in Fig.\,\ref{fig:bbr}(b) for \X\,$(19,0,1/2)$, where it can be seen that also ortho-\Htp states attain non zero population (only the $G=1/2$ component, as we have $\Delta G=0$). The forbidden transitions lead to population in \X\,$(17,1)$, \X\,$(18,1)$, and \X\,$(19,1)$, and eventually through absorption and emission via the \A$(0,0/2)$ states also to some population in \X\,$(18,3)$ and \X\,$(17,3)$. 
Despite the effect of the $g/u$-mixing, the decay of the initial state is not significantly affected, because the line strengths of the forbidden transitions are too small to compete with direct photodissociation. The maximum population in the intermediately populated ortho-states remains two orders of magnitude smaller than the maximum population acquired by the intermediate para-states.

Fig.\,\ref{fig:bbr}(c) and (d) consider the state redistribution with and without g/u-mixing starting in the \X\,$(19,1)$ level.
The overall behavior is similar to the one discussed for \X\,$(19,0)$, with the main difference that the initial state with $N=1$ has now two bound states to transfer to, \A\,$(0,0)$ and \A\,$(0,2)$. It can be seen that the states \X\,$(18,1)$ and \X\,$(17,1)$ attain a larger population than the states \X\,$(18,3)$ and \X\,$(17,3)$, as the latter states can be only reached from level \A\,$(0,2)$.

The rovibronic state evolution at 77~K is presented in Fig.\,\ref{fig:bbr_77K}.
Most results are qualitatively the same as for room temperature; however, two main differences can be identified: (i.) Photodissociation rates are reduced compared to room temperature so that the bound-bound redistribution becomes more important and intermediate levels reach around 3\% of the initial population. 
(ii.) The vibrational state with the largest acquired population is \X\,$(17)$ instead of \X\,$(18)$. This is because the population transfer between \A\,$(0)$ and \X\,$(18)$ at 293~K is mainly driven by stimulated emission. However, at 77~K, this rate of stimulated emission is about 33\% lower and the redistribution is mostly driven by spontaneous emission, which favors population transfer to the \X\,$(17)$ level.

The lifetime of a quantum state can be calculated from the spontaneous emission rate as
\begin{align}
    1/\tau_j
    & = \sum_i A_{ij} \, \text{.}
\end{align}
\noindent In the presence of resonant radiation, an effective lifetime can be defined that also takes into account stimulated emission and absorption \cite{demtroder2014a}:
\begin{align}
    \label{eq:tau_eff}
    1/\tau_{\text{eff},j}
    & = \sum_i \left[ A_{ij} + \bar B_{ij} \right] + \bar B_{\text{cont},j}  \, \text{.}
\end{align}

\begin{figure}
    \includegraphics[scale=1]{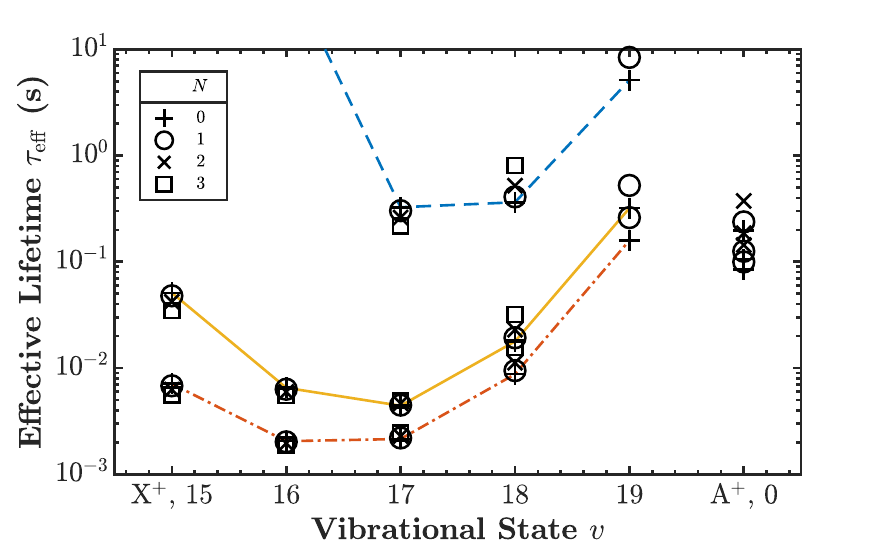}
    \caption{
    Effective lifetimes of the weakly bound states of \Htp.
    A line is drawn connecting states with $N=0$; dashed blue line for 77~K, solid orange line for 293~K and dash-dotted red line for 400~K.
    }
    \label{fig:lifetime}
\end{figure}

These effective lifetime were calculated for all weakly bound states by solving the rate equations and determining the time at which the normalized population of the initial state reached $1/e$. 
Estimation of the effective lifetimes were obtained using \citeeq{eq:tau_eff} and were found to agree with the numerical results within 3\%.
The results for 293~K and 77~K are presented in Table \ref{table:tau} and \ref{table:tau_77K}, respectively, and are displayed in Fig.\,\ref{fig:lifetime}. The following observations can be made:
(i.) At 293~K, the lifetimes range from 4~ms for \X\,$(17,0)$ to 523~ms for \X\,$(19,1)$, whereas at 77~K, they range from 216~ms for \X\,$(17,0)$ -- two orders of magnitude more -- to 1~min for \X\,$(16,0)$ or days for \X\,$(15,0)$.
(ii.) The lifetimes of the various rotational levels for a particular vibrational state are very similar. (iii.) The lifetimes of the \A\, states are not significantly affected by the change in temperature. (iv.) The effect of the g/u-mixing is insignificant.

It should be noticed that for states with $\bar B_{\text{cont}}>>\bar B\sim A$, the effective lifetime from \citeeq{eq:tau_eff} becomes $\tau_{\text{eff}} \approx 1/\bar B_{\text{cont}}$, so absorption to the continuum is the determining factor for the lifetime.

\begin{table}[!htb]
    \caption{
    Effective lifetimes of the weakly bound states of \Htp\, at 293~K.
    Level $v=0$ corresponds to \Ap\,, the other levels to \Xp.
    Table entries $a[b]$ stand for $a\times 10^b$.
    }
    \centering
    \begin{tabular*}{0.46\textwidth}{ c d{0.4} d{4.4} d{6.4} d{6.4} r }
        \multicolumn{6}{c}{$\tau_{\text{eff}}(293\text{~K})$ (s)} \\
        \\[-1em]
        \hline
        \\[-1em]
        \multicolumn{1}{c}{$v$} & \multicolumn{1}{c}{~~~~~ $N=0$} & \multicolumn{1}{r}{~~~ 1 ~~~} & 
            \multicolumn{1}{r}{~~~ 2 ~~~} & \multicolumn{1}{r}{~~~ 3 ~~~} & \\ 
        \\[-1em]
        \hline\hline
        \\[-1em]
             0 &  1.0620[-01] &  1.2558[-01] &  1.8600[-01] &  ~           & ~~~ \\
            19 &  3.1955[-01] &  5.2335[-01] &  ~           &  ~           & ~~~ \\
            18 &  1.7810[-02] &  1.9270[-02] &  2.2940[-02] &  3.2010[-02] & ~~~ \\
            17 &  4.4200[-03] &  4.4800[-03] &  4.6400[-03] &  4.9300[-03] & ~~~ \\
            16 &  6.5200[-03] &  6.3200[-03] &  5.9600[-03] &  5.4900[-03] & ~~~ \\
            15 &  5.1070[-02] &  4.7850[-02] &  4.2070[-02] &  3.4890[-02] & ~~~ 
    \end{tabular*}
    \label{table:tau}
\end{table}
\begin{table}[!htb]
    \caption{
    Effective lifetimes of the weakly bound states of \Htp\, at 77~K.
    Level $v=0$ corresponds to \Ap\,, the other levels to \Xp.
    Table entries $a[b]$ stand for $a\times 10^b$.
    }
    \centering
    \begin{tabular*}{0.46\textwidth}{ c d{0.4} d{4.4} d{6.4} d{6.4} r }
        \multicolumn{6}{c}{$\tau_{\text{eff}}(77\text{~K})$ (s)} \\
        \\[-1em]
        \hline
        \\[-1em]
        \multicolumn{1}{c}{$v$} & \multicolumn{1}{c}{~~~~~ $N=0$} & \multicolumn{1}{r}{~~~ 1 ~~~} & 
            \multicolumn{1}{r}{~~~ 2 ~~~} & \multicolumn{1}{r}{~~~ 3 ~~~} & \\ 
        \\[-1em]
        \hline\hline
        \\[-1em]
             0 &  1.9700[-01] &  2.3700[-01] &  3.7400[-01] &  ~           & ~~~ \\
            19 &  5.1190[+00] &  8.3910[+00] &  ~           &  ~           & ~~~ \\
            18 &  3.6300[-01] &  4.0900[-01] &  5.2000[-01] &  8.0700[-01] & ~~~ \\
            17 &  3.2600[-01] &  3.0300[-01] &  2.6100[-01] &  2.1600[-01] & ~~~ \\
            16 &  5.8191[+01] &  4.8358[+01] &  3.3474[+01] &  1.9685[+01] & ~~~ \\
            15 &  8.9116[+04] &  1.9420[+05] &  3.3290[+05] &  4.3669[+05] & ~~~ 
    \end{tabular*}
    \label{table:tau_77K}
\end{table}
\begin{table}[!htb]
    \caption{
    Effective lifetimes of the weakly bound states of \Htp\, at 400~K.
    Level $v=0$ corresponds to \Ap\,, the other levels to \Xp.
    Table entries $a[b]$ stand for $a\times 10^b$.
    }
    \centering
    \begin{tabular*}{0.46\textwidth}{ c d{0.4} d{4.4} d{6.4} d{6.4} r }
        \multicolumn{6}{c}{$\tau_{\text{eff}}(400\text{~K})$ (s)} \\
        \\[-1em]
        \hline
        \\[-1em]
        \multicolumn{1}{c}{$v$} & \multicolumn{1}{c}{~~~~~ $N=0$} & \multicolumn{1}{r}{~~~ 1 ~~~} & 
            \multicolumn{1}{r}{~~~ 2 ~~~} & \multicolumn{1}{r}{~~~ 3 ~~~} & \\ 
        \\[-1em]
        \hline\hline
        \\[-1em]
             0 &  8.4630[-02] &  9.9690[-02] &  1.4399[-01] &  ~           & ~~~ \\
            19 &  1.5871[-01] &  2.6052[-01] &  ~           &  ~           & ~~~ \\
            18 &  8.7600[-03] &  9.4600[-03] &  1.1230[-02] &  1.5650[-02] & ~~~ \\
            17 &  2.1500[-03] &  2.2000[-03] &  2.2900[-03] &  2.4600[-03] & ~~~ \\
            16 &  2.0500[-03] &  2.0200[-03] &  1.9600[-03] &  1.8900[-03] & ~~~ \\
            15 &  7.1100[-03] &  6.8100[-03] &  6.2700[-03] &  5.5500[-03] & ~~~ 
    \end{tabular*}
    \label{table:tau_400K}
\end{table}

To estimate the error in the rate constants and the calculated effective lifetimes, we determine the sensitivity of $A$ and $\tau_{\text{eff}}$ regarding the following approximations:
(i) Using BO vibrational wave functions and energies results in a change of $\Delta A/ A = 0.0253$ and $\Delta \tau_{\text{eff}}/\tau_{\text{eff}} = 0.0191$.
(ii) Using adiabatic wave functions and energies leads to $\Delta A/ A = 0.0104$ and $\Delta \tau_{\text{eff}}/\tau_{\text{eff}} = 0.0012$.
(iii) Using adiabatic wave functions and nonadiabatic energies including relativistic and radiative corrections from \citeref{moss1993a}, but increasing the stepsize of the potential to $0.02a_0$, we obtain $\Delta A/ A = 0.0002$ and $\Delta \tau_{\text{eff}}/\tau_{\text{eff}} = 0.0016$.

\section{Discussion and Outlook}
\label{sec:discussion}

This work sought to calculate the rovibronic population redistribution induced by BBR to determine the viability of conducting microwave spectroscopy of the weakly bound states of state-selectively produced, trapped \Htp\, ions. The necessity to include the effect of BBR for estimating effective lifetimes is motivated by the overlap of the BBR spectrum at room temperature with the \Ap-\Xp\, electronic spectrum of \Htp. 

We presented the time evolution of the internal state distribution for room temperature and liquid nitrogen temperature.
Interestingly, it was found that the largest contribution to the population redistribution comes from absorption to the continuum, given by $\bar B_{\text{cont}}$.
These coefficients are one to two orders of magnitude larger than $A$ and $\bar B$ at 293~K, and of the same order of magnitude at 77~K.
The rovibrational level with the largest photodissociation rate is \X$(17,0)$ with $\bar B_{\text{cont}} = 224.504\,\text{~s}^{-1}$ at 293~K and, in contrast, $A = 0.715\,\text{~s}^{-1}$ and $\bar B = 0.638\,\text{~s}^{-1}$ for \A\,$(0,1)\rightarrow\,\,$\X$(17,0)$.

As the main result of this work, we find effective lifetimes of the order of milliseconds at 293~K, which can be increase to hundred of milliseconds or even seconds at 77~K. This finding is in agreement with the measurements reported in \citerefs{carrington1989a,carrington1989b,carrington1993b,carrington1996a}, which demonstrated the possibility of conducting molecular ion-beam experiments in the microwave range with an interaction time of 0.3~$\mu$s.

More importantly, the found lifetimes of the highly-excited, weakly bound states will allow for state-of-the art ion trap experiments with millisecond interaction times, similar to the ones carried out using the vibrational ground state in HD$^+$. Specifically, the lifetimes are sufficiently long to enable transport and trap-loading of the ions, prepared in a separate photoexciation region.  
The \X(v=19) levels are identified as optimal "transfer states" by having the longest lifetimes among the weakly bound states at room and liquid nitrogen temperature. Once the ions are moved into the trap, the population could be transferred to different science states by adiabatic passage using microwave radiation.
We also determined effective lifetimes at 400~K, which are shown in Fig.~\ref{fig:lifetime} and listed in Table~\ref{table:tau_400K}, for the case of elevated temperatures in the trap region, which might be caused by rf heating. Compared to room temperature, the lifetimes are reduced by about a factor of two.

The effect of $g/u$-mixing experienced mainly by the \X\,$(19,0)$ and \X\,$(19,1)$ levels was found to reduce their fluorescence lifetime by 30\% to about 1000~s \citeref{moss1993b}. In this work, we found that the effect of $g/u$-mixing is negligible when the interaction with BBR is included: the effective lifetime of \X\,$(19,1)$ is reduced by only 210~$\mu$s, about 0.011\%, and the effective lifetime of \X\,$(19,0)$ is reduced by merely 60~$\mu$s, about 0.001\%.

Other compelling findings of this study are the following:
(i)     The lifetime of \A\,$(0)$ does not significantly vary with temperature and barely doubles when going from 293~K to 77~K.
This is because the state has a vanishing PDCS well before the peak of the BBS and the contribution from dissociation mostly comes from the GHz region, where the BBS does not decrease as drastically with temperature as in the THz region (see Fig.\,\ref{fig:spectra}).
This may set a limit on the interaction time of experiments relying on this state;
(ii)    The states of \X\,$(\leq 18)$ cannot decay via electric dipole transitions and their lifetime from electric quadrupole transitions are on the order of days, nonetheless, as a consequence of BBR, they have an effective lifetime even shorter than the higher vibrational levels;
(iii)   For an initial population of $10^3$ ions in \X\,$(19)$ levels (see Fig.\,\ref{fig:bbr}), only one ion is expected to reach \A\,$(0)$ by the time this state reaches its maximum population, while most of the total population is transferred to the continuum. This means, that the high purity of state-selectively prepared molecular ion ensembles is conserved, enabling a high signal-to-noise ratio.

Experiments with trapped \Htp\, ions relying on weaker two-photon or electric-quadrupole transitions can make use of more strongly bound \X~states, for which BBR induced photodissociation is irrelevant. Several of such studies are currently pursued, in one case \Htp\, ions are produced in the vibrational ground state \X$(v=0,N)$ by photoionization inside a rf trap \cite{schmidt2020a}, whereas in the other case a single \Htp\, ion is injected into a Penning trap, which allows to determine the rovibrational state before driving a spectroscopic transition \cite{tu2021a}.

\begin{acknowledgments}
\noindent MB acknowledges NWO for a VENI grant (VI.Veni.202.140).
\end{acknowledgments}


\appendix

\section{Coupling constant for the Fermi contact interaction}
\label{sec:bF}

The Fermi coupling constants of the weakly bound states were obtained by averaging the electron density at the position of the nuclei using the rovibrational wave functions obtained in Section \ref{subsec:rovib}, that is,
\begin{align}
    b_F(\eta,v,N) = \frac{8\pi}{3} \frac{g_Sg_p\mu_B\mu_N}{4\pi\epsilon_0c^2} &\, 
    \bra{\eta v N} \rho_e(R) \ket{\eta v N} \,\text{,}
\end{align}
\noindent where
\begin{align}
    \rho_e(R) = \bra{\eta} \delta(\bm r - \bm R) \ket{\eta} \,\text{,}
\end{align}
\noindent which was taken from \citeref{beyer2018a}, and $\eta$ indicating the electronic state.
The results are listed in Tables \ref{table:bF} and \ref{table:bF_A}.

\begin{table}[h!]
    \caption{Effective coupling constants of the Fermi interaction for the weakly bound states of \Htp\, in \Xp.}
    \centering
    \begin{tabular}{ c c c }
        \multicolumn{3}{c}{\normalsize $b_F$ (MHz)} \\
        \\[-1em]
        \hline
        \\[-1em]
        $v$ & ~~~$N=1$~~~ & ~~~$3$~~~ \\
        \\[-1em]
        \hline\hline
        \\[-1em]
            13  &   728.41  &   726.70  \\
            14  &   722.46  &   721.02  \\
            15  &   717.73  &   716.58  \\
            16  &   714.25  &   713.41  \\
            17  &   712.07  &   711.58  \\
            18  &   711.27  &   711.17  \\
            19  &   711.38  &   
    \end{tabular}
    \label{table:bF}
\end{table}

\begin{table}[h!]
    \caption{Effective coupling constants of the Fermi interaction for the weakly bound states of \Htp\, in \Ap.}
    \centering
    \begin{tabular}{ c c c }
        \multicolumn{3}{c}{\normalsize $b_F$ (MHz)} \\
        \\[-1em]
        \hline
        \\[-1em]
        $v$ & ~~~$N=0$~~~ & ~~~$2$~~~ \\
        \\[-1em]
        \hline\hline
        \\[-1em]
            0   &   711.32  &   711.32
    \end{tabular}
    \label{table:bF_A}
\end{table}

\section{Numerical results from the electric dipole moment}
\label{sec:num_res}

The numerical results for the bound-bound line strengths and the Einstein rate coefficients computed in Section \ref{sec:theory} are summarized in Tables \ref{table:mue_para}-\ref{table:mue_ortho0}. Absorption rate coefficients for the excitation into the continuum are presented in Tables \ref{table:Bk} and \ref{table:Bk_77K}.

\begin{table*}[h!]
    \caption{
    Bound-bound line strengths and Einstein rate coefficients of the weakly bound states of \Htp\, for transitions involving \A\,$(0,1)$.
    Rate coefficients for absorption can be obtained using the tabulated values and \citeeq{eq:B_abs}.
    The first and second column of $\bar B$ present the values at 293~K and 77~K, respectively.
    * Level \X\,$(19)$ lies above \A\,$(0)$ and thus the rate coefficients, $A$ and $B$, reported in this row take the former as the upper state.
    $a[b]$ stands for $a\times 10^b$.
    }
    \centering
    \begin{tabular*}{0.75\textwidth}{ r c@{\extracolsep{\fill}} d{0.4} d{0.4} d{0.4} d{0.4} d{3.4} d{4.4} r }
        \multicolumn{9}{c}{ \Ap $\,\,\, v=0,N=1$ } \\
        \toprule
        \\[-1em]
        \multicolumn{1}{r}{$v$} & \multicolumn{1}{r}{$N$} & \multicolumn{1}{r}{~~~$\nu$ (THz)} & 
        \multicolumn{1}{r}{~~~$\mu_e~(e a_0)$} & \multicolumn{1}{r}{~~~~$S~(e^2 a_0^2)$} &
        \multicolumn{1}{r}{~~~$A$~(s$^{-1}$)} & \multicolumn{2}{c}{~~~$\bar B$~(s$^{-1}$)} & \\
        \\[-1em]
        \hline
           *19 & 0 &  5.2890[-02] &  7.8405[+00] &  1.2295[+02] &  6.8403[-04] &  7.8613[-02] &  2.0409[-02] & ~~~ \\
            18 & 2 &  3.5197[-01] &  4.7023[+00] &  8.8448[+01] &  4.8334[-02] &  8.1445[-01] &  1.9704[-01] & ~~~ \\
            18 & 0 &  6.4562[-01] &  3.7955[+00] &  2.8812[+01] &  9.7175[-02] &  8.7118[-01] &  1.9615[-01] & ~~~ \\
            17 & 2 &  3.8904[+00] &  6.4155[-01] &  1.6463[+00] &  1.2150[+00] &  1.3632[+00] &  1.1796[-01] & ~~~ \\
            17 & 0 &  4.5903[+00] &  5.4325[-01] &  5.9025[-01] &  7.1549[-01] &  6.3828[-01] &  4.3417[-02] & ~~~ \\
            16 & 2 &  1.1809[+01] &  9.7880[-02] &  3.8320[-02] &  7.9082[-01] &  1.3361[-01] &  5.0338[-04] & ~~~ \\
            16 & 0 &  1.2872[+01] &  8.5060[-02] &  1.4470[-02] &  3.8679[-01] &  5.3463[-02] &  1.2688[-04] & ~~~ \\
            15 & 2 &  2.3980[+01] &  1.6470[-02] &  1.0900[-03] &  1.8762[-01] &  3.7676[-03] &  6.0557[-08] & ~~~ \\
            15 & 0 &  2.5368[+01] &  1.4500[-02] &  4.2000[-04] &  8.6064[-02] &  1.3712[-03] &  1.1694[-08] & ~~~ 
    \end{tabular*}
    \label{table:mue_para}
\end{table*}
\begin{table*}[h!]
    \caption{
    Bound-bound line strengths and Einstein rate coefficients of the weakly bound states of \Htp\, for transitions involving \A\,$(0,2)$.
    See caption of Table \ref{table:mue_para} for details.
    Line strengths are given for states with $G=1/2$; the corresponding value for states with $G=3/2$ can be obtained according to \citeeq{eq:Sij} and are twice as large.
    }
    \centering
    \begin{tabular*}{0.75\textwidth}{ r c@{\extracolsep{\fill}} d{0.4} d{0.4} d{0.4} d{0.4} d{2.4} d{4.4} r }
        \multicolumn{9}{c}{ \Ap $\,\,\, v=0,N=2$ } \\
        \toprule
        \\[-1em]
        \multicolumn{1}{r}{$v$} & \multicolumn{1}{r}{$N$} & \multicolumn{1}{r}{~~~$\nu$ (THz)} & 
        \multicolumn{1}{r}{~~~$\mu_e~(e a_0)$} & \multicolumn{1}{r}{~~~~$S~(e^2 a_0^2)$} &
        \multicolumn{1}{r}{~~~$A$~(s$^{-1}$)} & \multicolumn{2}{c}{~~~$\bar B$~(s$^{-1}$)} & \\
        \\[-1em]
        \hline
           *19 & 1 &  1.7610[-02] &  9.1888[+00] &  3.3774[+02] &  2.3115[-02] &  8.0022[+00] &  2.0945[+00] & ~~~ \\
            18 & 3 &  1.5664[-01] &  5.5070[+00] &  1.8196[+02] &  5.2584[+00] &  2.0234[+02] &  5.1276[+01] & ~~~ \\
            18 & 1 &  5.9369[-01] &  3.5779[+00] &  5.1205[+01] &  8.0575[+01] &  7.8894[+02] &  1.7994[+02] & ~~~ \\
            17 & 3 &  3.2774[+00] &  6.2199[-01] &  2.3212[+00] &  6.1447[+02] &  8.6476[+02] &  9.1554[+01] & ~~~ \\
            17 & 1 &  4.4040[+00] &  4.6558[-01] &  8.6704[-01] &  5.5692[+02] &  5.2677[+02] &  3.8241[+01] & ~~~ \\
            16 & 3 &  1.0827[+01] &  8.9435[-02] &  4.7992[-02] &  4.5809[+02] &  9.3651[+01] &  5.3778[-01] & ~~~ \\
            16 & 1 &  1.2565[+01] &  7.0481[-02] &  1.9870[-02] &  2.9641[+02] &  4.3391[+01] &  1.1773[-01] & ~~~ \\
            15 & 3 &  2.2672[+01] &  1.4634[-02] &  1.2849[-03] &  1.1260[+02] &  2.8149[+00] &  8.2142[-05] & ~~~ \\
            15 & 1 &  2.4953[+01] &  1.1808[-02] &  5.5775[-04] &  6.5168[+01] &  1.1125[+00] &  1.1468[-05] & ~~~ 
    \end{tabular*}
    \label{table:mue_ortho2}
\end{table*}
\begin{table*}[h!]
    \caption{
    Bound-bound line strengths and Einstein rate coefficients of the weakly bound states of \Htp\, for transitions involving \A\,$(0,0)$.
    See caption of Table \ref{table:mue_para} for details.
    Line strengths are given for states with $G=1/2$; the corresponding value for states with $G=3/2$ can be obtained according to \citeeq{eq:Sij} and are twice as large.
    }
    \centering
    \begin{tabular*}{0.75\textwidth}{ r c@{\extracolsep{\fill}} d{0.4} d{0.4} d{0.4} d{0.4} d{2.4} d{4.4} r }
        \multicolumn{9}{c}{ \Ap $\,\,\, v=0,N=0$ } \\
        \toprule
        \\[-1em]
        \multicolumn{1}{r}{$v$} & \multicolumn{1}{r}{$N$} & \multicolumn{1}{r}{~~~$\nu$ (THz)} & 
        \multicolumn{1}{r}{~~~$\mu_e~(e a_0)$} & \multicolumn{1}{r}{~~~~$S~(e^2 a_0^2)$} &
        \multicolumn{1}{r}{~~~$A$~(s$^{-1}$)} & \multicolumn{2}{c}{~~~$\bar B$~(s$^{-1}$)} & \\
        \\[-1em]
        \hline
           *19 & 1 &  9.6431[-02] &  6.2200[+00] &  7.7376[+01] &  8.6959[-01] &  5.4620[+01] &  1.4038[+01] & ~~~ \\
            18 & 1 &  5.1487[-01] &  4.2445[+00] &  3.6032[+01] &  1.8491[+02] &  2.1014[+03] &  4.8869[+02] & ~~~ \\
            17 & 1 &  4.3252[+00] &  6.2061[-01] &  7.7032[-01] &  2.3435[+03] &  2.2734[+03] &  1.6960[+02] & ~~~ \\
            16 & 1 &  1.2486[+01] &  9.7541[-02] &  1.9029[-02] &  1.3927[+03] &  2.0692[+02] &  5.8105[-01] & ~~~ \\
            15 & 1 &  2.4875[+01] &  1.6667[-02] &  5.5558[-04] &  3.2150[+02] &  5.5611[+00] &  5.9428[-05] & ~~~ 
    \end{tabular*}
    \label{table:mue_ortho0}
\end{table*}

\begin{table}[]
    \caption{Continuum absorption rates of the weakly bound states of \Htp\, at 293~K. $a[b]$ stands for $a\times 10^b$.}
    \centering
    \begin{tabular*}{0.46\textwidth}{ c d{0.4} d{4.4} d{6.4} d{6.4} r }
        \multicolumn{6}{c}{$\bar B_{\text{cont}}(293\text{~K})$ (s$^{-1}$)} \\ 
        \toprule
        \\[-1em]
        \multicolumn{1}{c}{$v$} & \multicolumn{1}{r}{~~~~~ $N=0$} & \multicolumn{1}{r}{~~~ 1 ~~~}  & 
        \multicolumn{1}{r}{~~~ 2 ~~~} & \multicolumn{1}{r}{~~~ 3 ~~~} & \\ 
        \\[-1em]
        \hline
        \\[-1em]
             0 &  4.1756[-01] &  5.6180[-01] &  6.6154[-01] &  ~           & ~~~ \\
            19 &  3.0503[+00] &  1.8482[+00] &  ~           &  ~           & ~~~ \\
            18 &  5.3561[+01] &  4.9919[+01] &  4.3118[+01] &  3.1106[+01] & ~~~ \\
            17 &  2.2450[+02] &  2.2149[+02] &  2.1496[+02] &  2.0232[+02] & ~~~ \\
            16 &  1.5328[+02] &  1.5815[+02] &  1.6787[+02] &  1.8211[+02] & ~~~ \\
            15 &  1.9577[+01] &  2.0902[+01] &  2.3772[+01] &  2.8667[+01] & ~~~ 
        \end{tabular*}
    \label{table:Bk}
\end{table}
\begin{table}[]
    \caption{Continuum absorption rates of the weakly bound states of \Htp\, at 77~K. $a[b]$ stands for $a\times 10^b$.}
    \centering
    \begin{tabular*}{0.46\textwidth}{ c d{0.4} d{4.4} d{6.4} d{6.4} r }
        \multicolumn{6}{c}{$\bar B_{\text{cont}}(77\text{~K})$ (s$^{-1}$)} \\ 
        \toprule
        \\[-1em]
        \multicolumn{1}{c}{$v$} & \multicolumn{1}{r}{~~~~~ $N=0$} & \multicolumn{1}{r}{~~~ 1 ~~~}  & 
        \multicolumn{1}{r}{~~~ 2 ~~~} & \multicolumn{1}{r}{~~~ 3 ~~~} & \\ 
        \\[-1em]
        \hline
        \\[-1em]
             0 &  9.7873[-02] &  1.3250[-01] &  1.2351[-01] &  ~           & ~~~ \\
            19 &  1.7430[-01] &  1.0228[-01] &  ~           &  ~           & ~~~ \\
            18 &  2.1915[+00] &  2.0166[+00] &  1.8094[+00] &  1.2036[+00] & ~~~ \\
            17 &  2.9510[+00] &  3.2089[+00] &  3.7731[+00] &  4.5669[+00] & ~~~ \\
            16 &  1.6840[-02] &  2.0366[-02] &  2.9630[-02] &  5.0481[-02] & ~~~ \\
            15 &  2.3250[-06] &  3.0429[-06] &  5.1857[-06] &  1.1280[-05] & ~~~  
        \end{tabular*}
    \label{table:Bk_77K}
\end{table}

\bibliography{bbr_H2p.bib}

\end{document}